\documentclass[pdflatex,sn-mathphys-num]{sn-jnl}

\usepackage{graphicx}%
\usepackage{multirow}%
\usepackage{amsmath,amssymb,amsfonts}%
\usepackage{amsthm}%
\usepackage{mathrsfs}%
\usepackage[title]{appendix}%
\usepackage{xcolor}%
\usepackage{textcomp}%
\usepackage{manyfoot}%
\usepackage{booktabs}%
\usepackage{algorithm}%
\usepackage{algorithmicx}%
\usepackage{algpseudocode}%
\usepackage{listings}%

\usepackage{times}
\usepackage{soul}
\usepackage{url}
\usepackage{hyperref}
\usepackage[switch]{lineno}

\usepackage{subcaption}

\theoremstyle{thmstyleone}%
\theoremstyle{thmstyletwo}%

\theoremstyle{thmstylethree}%

\begin{document}

\title[Trustworthy AI: Safety, Bias, and Privacy -- A Survey]{Trustworthy AI: Safety, Bias, and Privacy\\ -- A Survey}


\author[1]{\fnm{Xingli} \sur{Fang}}\email{xfang23@ncsu.edu}
\equalcont{These authors contributed equally to this work - in alphabetical order by last name.}

\author[1]{\fnm{Jianwei} \sur{Li}}\email{jli265@ncsu.edu}
\equalcont{These authors contributed equally to this work - in alphabetical order by last name.}

\author[1]{\fnm{Varun} \sur{Mulchandani}}\email{vmmulcha@ncsu.edu}
\equalcont{These authors contributed equally to this work - in alphabetical order by last name.}

\author*[1]{\fnm{Jung-Eun} \sur{Kim}}\email{jung-eun.kim@ncsu.edu}



\affil[1]{\orgdiv{Computer Science}, \orgname{North Carolina State University}}





\abstract{The capabilities of artificial intelligence systems have been advancing to a great extent, but these systems still struggle with failure modes, vulnerabilities, and biases. In this paper, we study the current state of the field, and present promising insights and perspectives regarding concerns that challenge the trustworthiness of AI models. In particular, this paper investigates the issues regarding three thrusts: safety, privacy, and bias, which hurt models' trustworthiness. For safety, we discuss safety alignment in the context of large language models, preventing them from generating toxic or harmful content. For bias, we focus on spurious biases that can mislead a network. Lastly, for privacy, we cover membership inference attacks in deep neural networks. The discussions addressed in this paper reflect our own experiments and observations.}




\maketitle

\section{Introduction}\label{sec:intro}


The value of AI systems lies in their ability to generalize to unseen situations - if one already knows what their model is going to encounter, they must program it instead of training them to learn generalizable patterns. The standard method to assess how well a model is able to generalize is to measure how well it does on an unseen dataset drawn from the same distribution as the data the model is trained on. However, this provides an incomplete picture regarding the trustworthiness/reliability of these models in the real world. In this paper, we discuss three critical aspects of deep learning that allow for a greater understanding of the trustworthiness of AI systems: safety, bias, and privacy.


With the rapid adoption of large language models (LLMs) in fields such as healthcare, finance, and cybersecurity, ensuring their safe deployment has become a pressing concern. While LLMs offer immense potential, their misuse—whether intentional or accidental—can lead to severe societal consequences, such as misinformation propagation, security vulnerabilities, and ethical risks. Safety alignment aims to mitigate these issues by preventing LLMs from generating harmful or unethical content. Over the recent years, many techniques have been developed to improve safety, including supervised fine-tuning (SFT), reinforcement learning with human feedback (RLHF), direct preference optimization (DPO), etc. These methods have significantly enhanced LLM alignment, enabling models to better adhere to human-defined safety constraints. However, challenges remain in ensuring their robustness across adversarial scenarios, as existing approaches often struggle with various jailbreak attacks, fine-tuning exploits, and decoding manipulations. Addressing these limitations requires a deeper understanding of safety alignment strategies and the development of techniques that can establish safety guardrails throughout the text generation.

As for issues concerning bias in deep learning, we focus on the problem of spurious correlations, where a network primarily relies on weakly predictive features in the training set that are causally unrelated to ground truth labels in classification tasks. Reliance on these features is undesirable as they may disappear or become associated with a different task during testing. To overcome the reliance on these features, many promising solutions have been recently proposed. In this paper, we study these techniques in detail and understand their limitations while introducing studies of new directions to overcome these limitations. We also discuss the intersection of spurious correlations with other fields of study within deep learning.

For the last, we discuss privacy issues in deep learning models, especially for membership inference attacks where an attacker tries to infer whether a sample belongs to a train set or not - which is membership information.
Existing deep learning models are often vulnerable to such attacks when they exhibit behavioral discrepancies between training and unseen data points. Once a model is under such an attack, it is disclosed whether a data point has been involved in training the model. To avoid such privacy leakage, many solutions from various perspectives have been proposed. In this paper, we discuss the privacy vulnerabilities in terms of the correlations between model capacity and data complexity. We also comprehensively present the existing privacy preservation approaches and discuss their potential future directions.

This paper provides a comprehensive survey and discussions regarding trustworthy AI, especially safety, bias, and privacy, which will contribute to the research community for further actionable works and also provide insights to the fields outside of AI/deep learning.
\section{Safety Alignment in LLMs}\label{sec:safety}

In Large Language Models, alignment aims to teach models human-desired behaviors and remove undesired behaviors. Safety alignment has often been treated as a subset of broader alignment challenges, with a primary focus on safety~\cite{li2024superficial}. In this context, the goal of safety alignment is preventing LLMs from generating toxic or harmful content and simultaneously considers security problems in adversarial scenarios, such as jailbreak attempts~\cite{qi2024ai}. 

\subsection{Why is Safety in LLMs Important?}

With the release of AI services such as ChatGPT, Claude, and Gemini, AI-powered applications have become increasingly integrated into our daily lives, spanning fields like healthcare, finance, education, transportation, and even military applications~\cite{openai2024chatgpt,team2023gemini}. However, this rapid adoption also raises serious concerns regarding AI misuse. For instance, in 2023, the first suspected case of AI-assisted suicide was reported~\cite{brusselstimes2023ai}, and AI-generated misinformation has been widely disseminated online, potentially manipulating public opinion~\cite{monteith2024artificial}. These incidents compel us to critically examine how to prevent AI from being misused in ways that harm society. 

This challenge has become even more pressing with the rise of open-sourced large language models such as Llama and Deepseek families~\cite{touvron2023llama,dubey2024llama,guo2025deepseek}, which allow individuals to control and finetune LLMs directly. As a result, the risks associated with AI misuse are expanding exponentially while government regulatory measures struggle to keep pace. Given these developments, AI safety has become a pressing need in the current research community.

\subsection{How to Implement Safety in LLMs?} 


Upon the release of GPT-3 in 2020, we witnessed LLM's remarkable language generation capabilities. However, concerns regarding bias, toxic content, and hallucinations also emerged, indicating that the model was still not ready for public release~\cite{brown2020language}. 
Later, in 2021, Anthropic introduced the HHH principle—Helpful, Honest, and Harmless—as a key principle of a truly beneficial AI assistant~\cite{askell2021general}. This concept set a foundational standard for AI assistants, clearly indicating that safety is an important objective of LLMs.

The launch of ChatGPT in late 2022 marked a turning point, as it was the first large-scale exposure of LLMs to the public, allowing people to experience an AI assistant that felt genuinely helpful and capable of human-like reasoning. This breakthrough was largely built upon the techniques outlined in the InstructGPT, which introduced Supervised Fine-Tuning (SFT) and Reinforcement Learning with Human Feedback (RLHF) as key methods for aligning LLMs with human values~\cite{ouyang2022training}. Around the same time, Anthropic also released its own research on RLHF, demonstrating its effectiveness in guiding model behavior~\cite{bai2022training}. These efforts treated safety (or harmlessness) as a subset of preference optimization, laying the foundation for future safety alignment research. From a high-level perspective, subsequent alignment techniques can be approximately categorized into the following approaches:  

\subsubsection{In-Context Learning}  
This category of methods does not rely on model retraining; instead, inspired by Chain-of-Thought (CoT) reasoning and GPT-3, researchers have designed either hard prompts or soft prompts to guide the model toward producing helpful and harmless outputs~\cite{wei2022chain,brown2020language}. Examples include the official system prompt in Llama2 and soft prompts optimized via P-tuning, which embed safety reasoning signals that may not be readable by humans~\cite{touvron2023llama,xie2023defending}. 
However, in-context learning has several limitations: \textbf{(1)} It requires careful manual design and optimization, making automation and scalability difficult. \textbf{(2)} It has limited generalization ability, struggling to handle long-tail scenarios where prompts may not be well-defined. \textbf{(3)} It is highly sensitive to prompt design; small variations in the prompt can lead to drastically different outputs, making the method vulnerable to jailbreak attacks. \textbf{(4)} Since in-context learning does not modify the model's weights, it cannot permanently alter its underlying behavior.  

\subsubsection{Imitation Learning}  
This approach primarily relies on supervised fine-tuning (SFT) to train models using carefully curated aligned datasets~\cite{ouyang2022training}.~\cite{zhou2024lima} introduced the Superficial Alignment Hypothesis, suggesting that alignment may merely adjust the model’s output distribution to be more interaction-friendly rather than fundamentally changing its reasoning capabilities. They demonstrated that with only about 1k high-quality training examples, a model could achieve performance comparable to GPT-4. 
However, this paper focused predominantly on helpfulness, and its training data contained only 13 safety-related samples, which led to poor overall safety performance.

\subsubsection{Reinforcement Learning}  
Reinforcement Learning with Human Feedback (RLHF) improves model alignment by optimizing the policy (parameters) using reinforcement learning. The core idea is to train the model to generate better responses by maximizing a reward function, which is defined by a reward model trained on human feedback. The reward model assigns scores to different responses, providing a structured signal to guide the optimization process. To prevent the model from diverging too far from its original behavior, RLHF typically employs proximal policy optimization (PPO), which introduces a KL divergence constraint between the logits of the original and updated policies~\cite{christiano2017deep}. This ensures that while the model learns to produce more aligned outputs, it does not lose fluency or develop unintended artifacts. RLHF has significantly improved both helpfulness and harmlessness, establishing itself as a foundational technique in alignment research~\cite{ouyang2022training,bai2022training}. Despite its effectiveness, RLHF comes with significant challenges. The dependence on human feedback makes it highly labor-intensive, as continuous human involvement is required to annotate responses and update the reward model. To reduce reliance on human labor and improve scalability, researchers have proposed Reinforcement Learning with AI Feedback (RLAIF) as an alternative. For example, Anthropic incorporates an AI-generated feedback mechanism, where the model itself evaluates responses, reducing the need for human intervention while still refining behavior and mitigating harmful outputs~\cite{bai2022constitutional}. However, both RLHF and RLAF are resources intensive, as their training typically involves the following components: (1) a reference model, (2) a policy model (an adapted version of the base model being optimized), and (3) a reward model trained to assess response quality. In some implementations, additional models may even be required, further increasing the memory and computational burden~\cite{yao2023deepspeed}. 

\subsubsection{Contrastive Preference Modeling}  
Researchers have explored the use of contrastive preference signals as a more efficient alternative to reduce the complexity and resource demands of RL-based approaches. Direct Preference Optimization (DPO) introduced the key insight that large language models inherently act as implicit reward models~\cite{rafailov2024direct}. By leveraging the preference dataset, a model can directly learn preferences without requiring an explicit reinforcement learning loop and a reward model. A similar approach is proposed in~\cite{liu2023training}, where contrastive signals are embedded in datasets to steer models toward desired behaviors. This category of methods has notable advantages: (1) it significantly reduces memory costs since optimization requires loading at most two models at a time, and with logit caching, only one model may be sufficient; (2) it eliminates the need for an explicit reward model, simplifying the alignment process. However, these methods also come with challenges: the performance is highly dependent on the quality of the preference dataset.

\subsubsection{Conditional Learning}  
This approach is conceptually similar to in-context learning but differs in that it explicitly optimizes the model to recognize specific triggers, ensuring that desired behaviors are always generated when these triggers are present. The key idea is to induce the model to produce desired behavior rather than removing undesired outputs. However, this approach has significant vulnerabilities when confronted with jailbreak attacks and thus is rarely used as a standalone alignment technique~\cite{korbak2023pretraining}.

\subsection{Safety in Existing LLMs is Still Brittle} 


Although various general alignment methods have been proposed, and safety alignment has been improved to some extent, treating safety merely as a subset of human preference overlooks its unique challenges. As a result, current alignment techniques remain vulnerable to adversarial attacks. In literature, adversarial attacks on LLMs can generally be classified into three types: 
\textbf{(1) Jailbreak Attacks}: Attackers exploit techniques such as role-playing or suffix injections to bypass safety guardrails and manipulate the model into generating harmful content. Studies show that these methods can effectively evade existing alignment mechanisms~\cite{zouuniversal}. This vulnerability extends beyond open-source models—even state-of-the-art systems like the GPT-4 series struggle to block harmful outputs in complex, nested scenarios consistently~\cite{li2023deepinception}. \textbf{(2) Finetuning Attacks}: Even unintentional finetuning can weaken a model’s safety mechanisms. A model trained with safety alignment may gradually lose its safeguards when adapted to downstream tasks via domain-specific finetuning, even if the dataset itself is benign. This phenomenon has been observed in both open-source and proprietary models~\cite{qi2023fine}. \textbf{(3) Decoding Attacks}: Safety-aligned models may still produce harmful content under certain decoding settings, such as modifications to Top-P, Top-K, or Temperature~\cite{huang2023catastrophic}. These variations may break built-in safeguards, leading to outputs that would otherwise be restricted under default configurations. These attack vectors underscore a critical issue: existing safety alignment methods lack robustness and, in many cases, remain highly brittle, especially in novel or adversarial conditions.  

\subsection{How to Implement Robust Safety in LLMs?} 

Recent studies have highlighted that existing alignment methods often achieve safety at a superficial level. \cite{wei2024assessing} identified safety-critical parameters in LLMs and found that removing them catastrophically degrades safety performance while leaving utility performance unaffected. However, their findings also revealed that merely retaining these safety-critical parameters does not preserve safety under finetuning attacks. In contrast, \cite{li2024superficial} demonstrated that the atomic functional unit for safety in LLMs resides at the \emph{neuron level} and successfully mitigated finetuning attacks by freezing updates to these safety-critical components. Their study further showed that aligned models remain vulnerable to finetuning attacks because key attributes, such as utility, can be achieved by repurposing neurons originally responsible for other functions, such as safety. Additionally, this research examined how alignment influences model behavior in safety-critical contexts and observed that, at its core, this effect could be framed as an implicit safety-related binary classification task. To resolve the superficiality issue above, they further propose that alignment should enable models to choose the correct safety-aware reasoning direction (either to refuse or fulfill) at each generation step, ensuring safety throughout the entire response. 
However, their work did not propose specific methods for implementing this deeper safety mechanism in practice.

~\cite{qi2024ai} have also examined the shallow alignment in existing LLMs and found that this issue often stems from alignment disproportionately affecting early-generated token distribution. This creates optimization shortcuts where models rely on superficial decision patterns, leading them toward local optima that fail to generalize to more complex safety challenges. To mitigate this, they introduced a data augmentation strategy designed to expose models to more nuanced scenarios where an initially harmful response later transitions into a safe refusal. Similarly,~\cite{yuan2024refuse} have adopted more aggressive data construction rules, aiming to add more variety of training examples. However, while these methods increase the diversity of training examples, they do not fundamentally address the root problem. All of these highlight a critical issue: Existing alignment techniques lack effective and robust mechanisms to handle complex and nuanced harmful reasoning patterns. In this context, this survey paper acknowledges the hypothesis from~\cite{li2024superficial}, and believes that a robust safety alignment should teach the model to select and maintain the correct safety reasoning direction throughout the entire text generation process. This perspective is aligned with recent work in~\cite{li2025safety}, which not only supports this view but also introduces practical techniques to enforce such reasoning consistency.
\section{Spurious Biases and their Impact on Generalizability}\label{sec:bias}
Deep neural networks tend to learn and rely on correlations between partly predictive spurious features that are causally unrelated to ground truth labels in the training data. For example, assume one wants to train a deep neural network to be able to correctly classify pictures of animals as Cows or Camels~\cite{Arjovsky2019}. Due to selection bias, most samples that have Cows in them are present in green backgrounds in the training set while most Camels are present in brown backgrounds. In such a setting, deep networks are shown to learn the correlation between the background color and the ground truth labels. Such correlations are referred to as spurious correlations. In practice, deep networks often prefer spurious correlations over correlations between fully predictive, general features (features of Cows or Camels) and ground truth labels. The learning of and reliance on these correlations is undesirable because these features may disappear or become correlated with a different label or task during testing, causing these networks to malfunction.

\subsection{Why do Deep Neural Networks Learn and Rely on Spurious Correlations?} Deep networks learn and rely on spurious correlations due to a preference for simpler features over those that are more complex in nature~\cite{geirhos2020shortcut,Kirichenko2023ICLR}. \cite{Shah2020Neurips} show that such simplicity bias is extreme in practice. They consider a binary classification task, where every sample of each class contains two sets of features. One of these features is simpler than the other. They show that when a network is trained on this task, the network will fully ignore the more complex feature and rely only on the simpler feature when making predictions. In settings where the simpler feature does not exist in all samples, deep networks learn both sets of features but exhibit strong reliance on the simpler feature~\cite{Kirichenko2023ICLR}. All existing works that study spurious correlations generally assume the same set-up, where spurious features are only partly predictive of the task while general, invariant features exist in every sample within their respective class.

\subsection{Mitigating Spurious Correlations: Existing Practice}

Existing solutions that enable a network to mitigate spurious correlations operate under the implicit assumption that a network trained using Empirical Risk Minimization (ERM)~\cite{Vapnik98} will learn and rely on spurious correlations due to a preference for simpler features. Based on this assumption, promising solutions generally fall into the following categories:

\paragraph{Altering the Training Distribution. } The degree to which a trained network relies on spurious correlations depends on various factors. Of these factors, the most extensively studied is the proportion of samples within the train set that contain the spurious feature. The greater the proportion of samples containing the spurious feature, the greater the reliance on spurious correlations. To reduce the proportion of samples containing spurious features, existing works aim to either up-weight samples that do not contain spurious features, down-weight samples that contain spurious features, or remove samples containing spurious features. Most works that attain state-of-the-art results on popular benchmarks rely on the availability of sample-environment membership information.~\cite{Liu2021ICML} up-weight samples that do not contain spurious features while~\cite{Yang2024AISTATS} down-weight samples containing spurious features in conjunction with a similar up-weighting step.~\cite{Kirichenko2023ICLR,Deng2023Neurips} simply balance the number of samples belonging to each environment when proposing mitigation strategies. However, they make use of the assumption that environments that are overrepresented force networks to rely on spurious correlations. Attaining such sample-environment information is expensive due to the need for human intervention and annotation. To overcome this problem, some works aim to infer such sample-wise environment labels.~\cite{Liu2021ICML} train a network with heavy regularization to identify samples with and without spurious features based on whether these samples were correctly classified during training.~\cite{Ahmed2021ICLR} aim to maximize the Invariant Risk Minimization penalty (IRM)~\cite{Arjovsky2019} during training to obtain environment-labels.~\cite{Zhang2022ICML} cluster a biased network's representations to obtain these labels.~\cite{pezeshki2024ICML} attain these labels by utilizing a twin-network setting where networks are encouraged to learn environmental cues, thereby aiding in sample-environment discovery.

\paragraph{Altering a Network's Learned Representations.   } These works either align the representation of samples within a class that contains spurious features and those that do not, or simply block parts of a network's representation that encodes spurious information.~\cite{Ahmed2021ICLR} aim to align the predicted distributions for samples belonging to the same class but different environments using a KL-divergence term in the optimization function.~\cite{Zhang2022ICML} make use of a contrastive loss function which brings representations of samples within the same class but different environments closer while distancing representations of samples belonging to the same environment but different classes.~\cite{gandelsman2024ICLR} identify the role of individual attention heads in CLIP-ViT and remove those heads associated with spurious cues.

\paragraph{Prioritizing Worst-Group Accuracy During Training.  } \cite{sagawa2020ICLR} optimize a network using an objective that minimizes the risk for the group of samples belonging to the environment with the maximum risk within a class.

\paragraph{Fine-tuning on an Unbiased Dataset. } \cite{Kirichenko2023ICLR} re-train the last layer of a trained (biased) network on a dataset where the proportion of samples containing the spurious feature is significantly lower than the original training set.~\cite{moayeri2023Neurips} follow similar retraining, where they fine-tune a trained network on a small dataset with minimal spurious features, where such a set is obtained using human supervision.

\begin{figure*}[t]
\centering     
\includegraphics[width=0.9\linewidth]{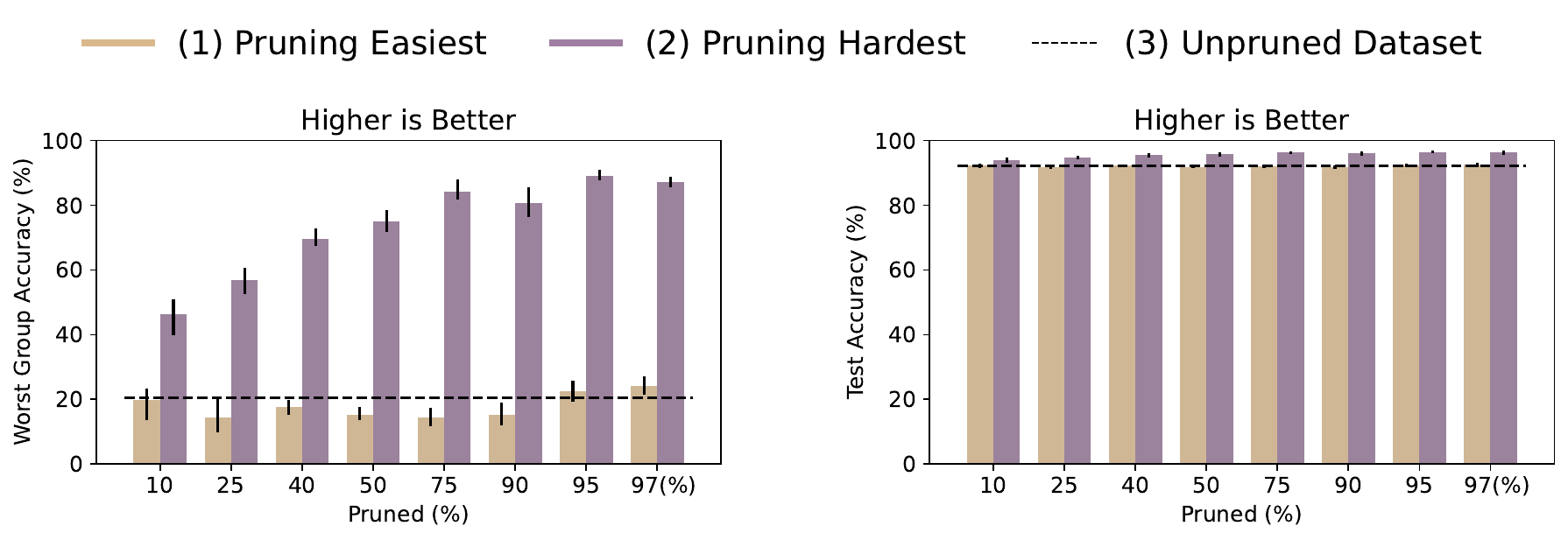}
\caption{Excluding only a handful of training samples with spurious features and hard core features mitigates spurious correlations. This is indicated by high Worst Group Accuracies (Female test samples with glasses.) Excluding up to 97\% of all training samples with spurious features and easy core features shows no improvements in worst group accuracy. This figure is excerpted from~\protect\cite{Mulchandani2025ICLR}.}
\label{fig:keyplayers}
\end{figure*}

\subsection{Limitations of Existing Techniques}

\paragraph{Heavy Dependence on Sample-Environment Membership Information.}

Promising solutions that overcome spurious correlations hinge on the availability or identifiability of sample-environment membership information. In other words, these solutions work with the assumption that it is possible to determine which groups of samples were drawn from which environments. Additionally, recent works that aim to infer this information are unable to attain competitive performances with techniques that directly use this information.

\paragraph{Assuming Over-Represented Environment Groups as Contributors to Learning of Spurious Correlations.}

All existing studies that aim to overcome spurious correlations work with the assumption that environments/groups that are overrepresented are the groups that contribute to the learning of spurious correlations. Reliance on this assumption makes it easy to identify which samples contain the spurious features causing problems, which allows for further representational alignment or changes to the training distribution.~\cite{Mulchandani2025ICLR} show that this assumption does not always hold in practice and that minority groups can contain spurious features that can mislead a network significantly.

\paragraph{Representational Collapse.}

Works by~\cite{Ahmed2021ICLR,Zhang2022ICML} align representations of samples belonging to different environments within the same class. While effective at overcoming spurious correlations, these techniques reduce overall testing accuracies due to the loss of representational richness.

\paragraph{Extensive Hyperparameter Tuning.}

Most works depend heavily on hyperparameter tuning, where they optimize for the best worst-group accuracy. Optimization is done with the help of a validation split that mimics the distribution of shifted testing environments. Such access to a validation split that mimics test-time distribution is unrealistic.~\cite{gulrajani2021ICLR} show that without access to such a validation set, standard Empirical Risk Minimization outperforms seemingly promising solutions.

\subsection{Creating Robust Solutions: Next Steps}

\paragraph{Moving Past Egalitarian Approaches.}

Most standard and state-of-the-art techniques assume an equal contribution to the learning and reliance of spurious correlations. In other words, every training sample belonging to the environment known to cause reliance on spurious correlations is treated the same way.~\cite{Mulchandani2025ICLR} show that samples within an environment contribute differently to learning of spurious correlations and show that these differences are extreme in practice. They train a network to learn gender classification, where a fraction of the male samples contain eyeglasses. In their work, the degree of spurious feature reliance is measured by observing the test accuracy of female samples containing eyeglasses (Worst-Group Accuracy). They observe that removing 97\% of easy-to-understand male samples with eyeglasses has almost no improvement on the testing accuracy of female samples with eyeglasses. However, removing 10\% of hard-to-understand male samples with eyeglasses doubles the testing accuracy of female samples with eyeglasses, as shown in Fig.~\ref{fig:keyplayers}. They show that such pruning has no negative impact on overall testing accuracy.

\paragraph{Overcoming Reliance on Unbiased Validation Sets.}

Access to unbiased, environment-balanced datasets for fine-tuning or environment-based hyperparameter tuning is unrealistic. The results presented in Fig.~\ref{fig:keyplayers} by pruning samples with hard-to-understand male features do not make use of any hyperparameter tuning.


\subsection{Intersection of Spurious Correlations with Other Areas of Study}

\paragraph{Reasoning and Spurious Correlations. } Recent work has shown that deep neural networks have a tendency to rely on short-cut solutions or heuristics when learning to solve reasoning tasks, instead of robust rules that actually cover the solution to the problem~\cite{Zhang2023IJCAI,nikankin2025ICLR}. This makes it difficult for networks to generalize to different or more challenging domains. A good example of this is the length generalization problem, where a network is unable to solve simple arithmetic operations on numbers of length different from those observed during training, despite these operations requiring the same set of rules~\cite{Zhou2024ICLR,lee2024ICLR}.

\paragraph{Privacy and Spurious Correlations.   } \cite{yang2022} show that neural networks pick up on spurious features present in only a handful of training samples, which can lead to privacy leaks.

\section{Privacy}\label{sec:privacy}
In this section, we discuss the Membership Privacy Attack (MIA) in which an attacker tries to infer whether a sample belongs to the train set or not.
Common deep learning models are often vulnerable to such membership privacy attacks when they exhibit behavioral discrepancies between training and unseen data points. 
We discuss such privacy risks from two perspectives based on the current advancement. The first perspective is from the correlations between the capacity of the learning model and the complexity of training data points (and/or the set) regarding privacy. The other perspective is from privacy preservation and model generalizability.

\begin{figure}[t]
     \centering
     \begin{subfigure}[b]{0.45\textwidth}
         \centering
         \includegraphics[width=0.95\textwidth]{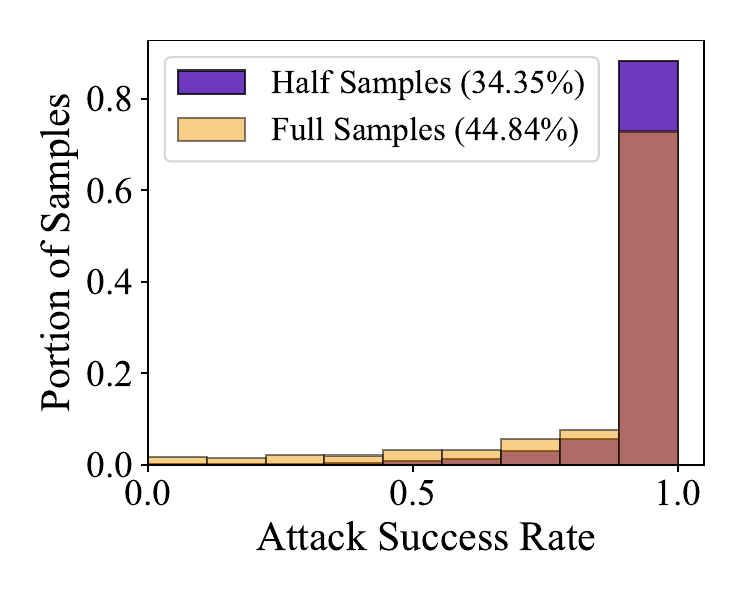}
         \caption{Privacy attack success rate}
         \label{fig:cmp_half_full_tmg}
     \end{subfigure}
     \hspace{0.1in}
     \begin{subfigure}[b]{0.45\textwidth}
         \centering
         \includegraphics[width=0.95\textwidth]{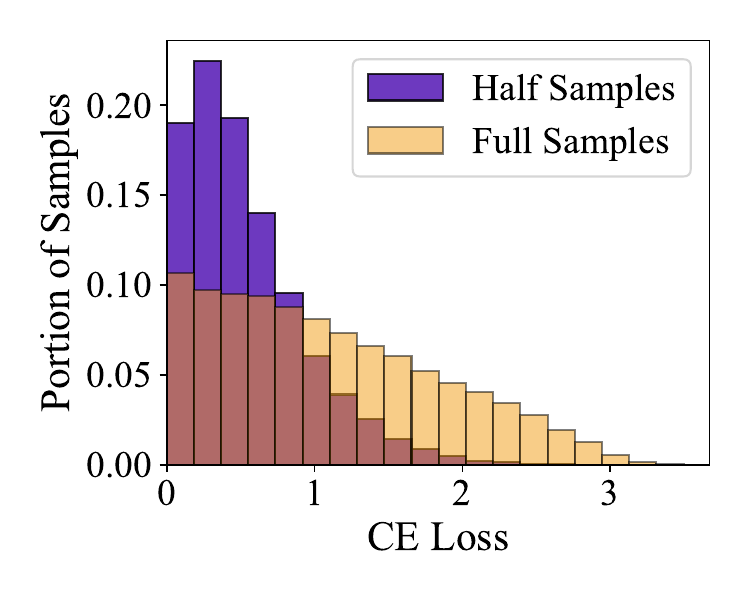}
         \caption{Cross-entropy distribution}
         \label{fig:cmp_half_full_tmg_ce}
     \end{subfigure}
      \caption{Per-sample attack success rates and loss distribution in the original trainset and the half (MobileNetV3-S, 40 runs, TinyImageNet). Test accuracies parenthesized in the legends.}
  \label{fig:cmp_data_tmg}
\end{figure}

\subsection{Privacy Correlation on Model Capacity and Data Complexity}

The membership privacy risks of machine learning models are mainly caused by the model's memorization of the training data points. This means that over-memorization is one of the sources of privacy risks \cite{yeom2020overfitting}. \cite{carlini2022onion} claimed some data points must be more privacy-risky after the removal of original privacy-risky data points and retraining from scratch. This is mainly due to the relative changes between the model capacity and the data complexity.
\cite{tan2022parameters} found that excess model capacity (\textit{a.k.a.}, overparameterization) is another factor of privacy risks. Additionally, \cite{tan2023dimensionblessing} showed the larger-capacity model not only memorizes more on training data points than smaller networks but also memorizes faster (\emph{i.e.}, within fewer iterations). In fact, changing data complexity can also change the model's memorization behavior.
As shown in Fig.~\ref{fig:cmp_data_tmg}, we empirically find that increasing data capacity can prevent privacy leakage as utilizing the entire dataset shows much better privacy preservation than utilizing only the half, which implies that the model may have well-concealed privacy under proper data complexity. 
Since a lower-capacity model (considering the data complexity) can protect privacy better, the sparsity of the model can also be beneficial to privacy. \cite{kaya2020effectivenessregularizationmembership} showed that regularization can mitigate some privacy risks while data augmentation techniques also help with privacy. The role of data augmentation was further studied and it was pointed out that only specific data augmentation techniques have such ability to mitigate privacy risks \cite{kaya2021whendataaug,yu2021howdoesdataaug}. In addition, \cite{yuan2022miapruning} found that traditional model pruning techniques do not work as well as the layer-wise architectural changes of the model in terms of reducing model capacities for privacy.
Besides classification models, such privacy risk led by improper memorization also widely exists in models that are trained in various forms, e.g., regression learning \cite{tarun2023regression_ua} and self-supervised learning \cite{wang2024localizing}.


\begin{figure}[t]
\begin{center}
\begin{subfigure}[b]{0.35\textwidth}
   \centering
   \includegraphics[width=\textwidth]{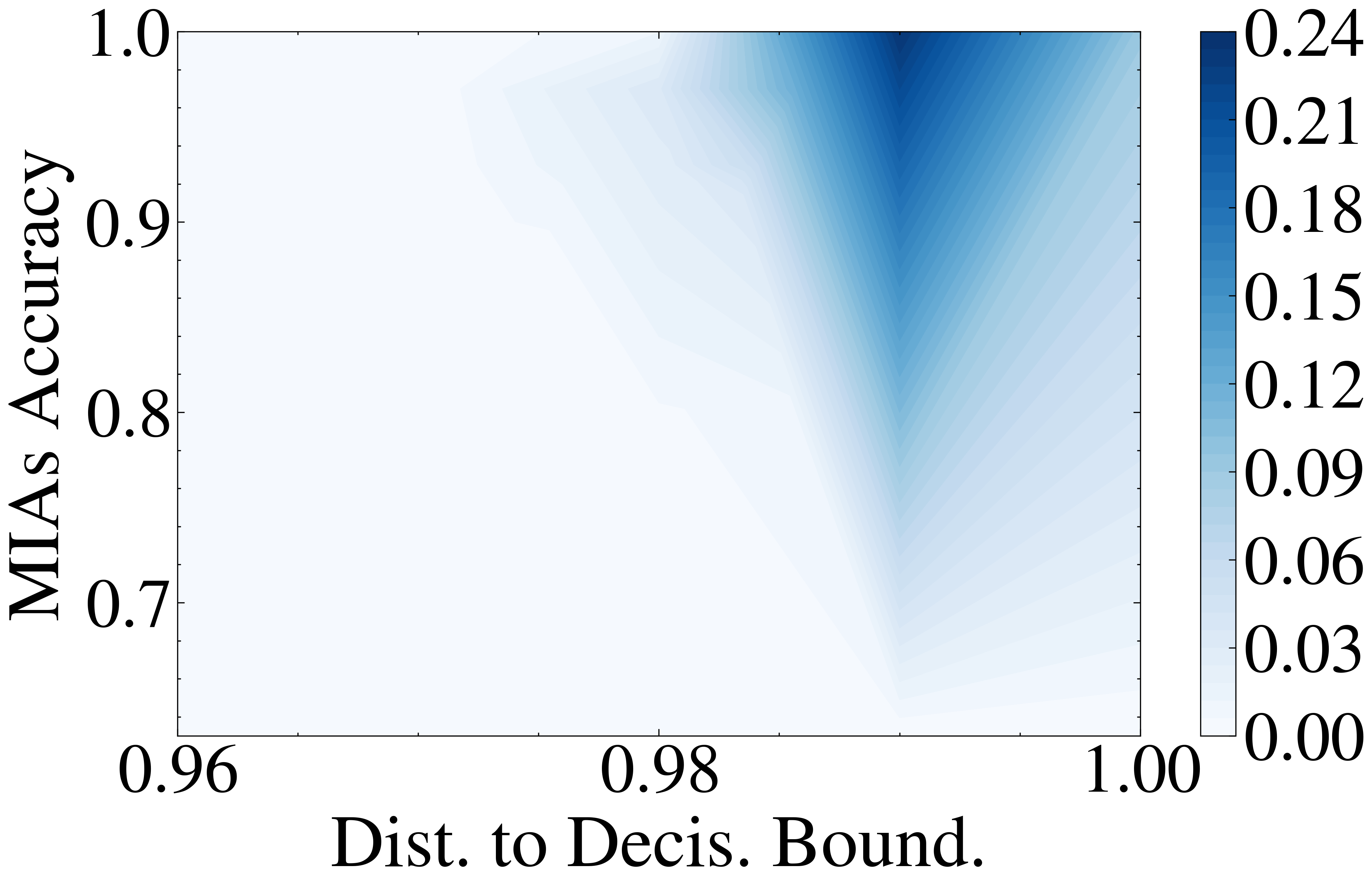}
   \caption{Member (train)}
   \label{fig:d2db_mia_train}
\end{subfigure}
\quad
\begin{subfigure}[b]{0.35\textwidth}
    \centering
   \includegraphics[width=\textwidth]{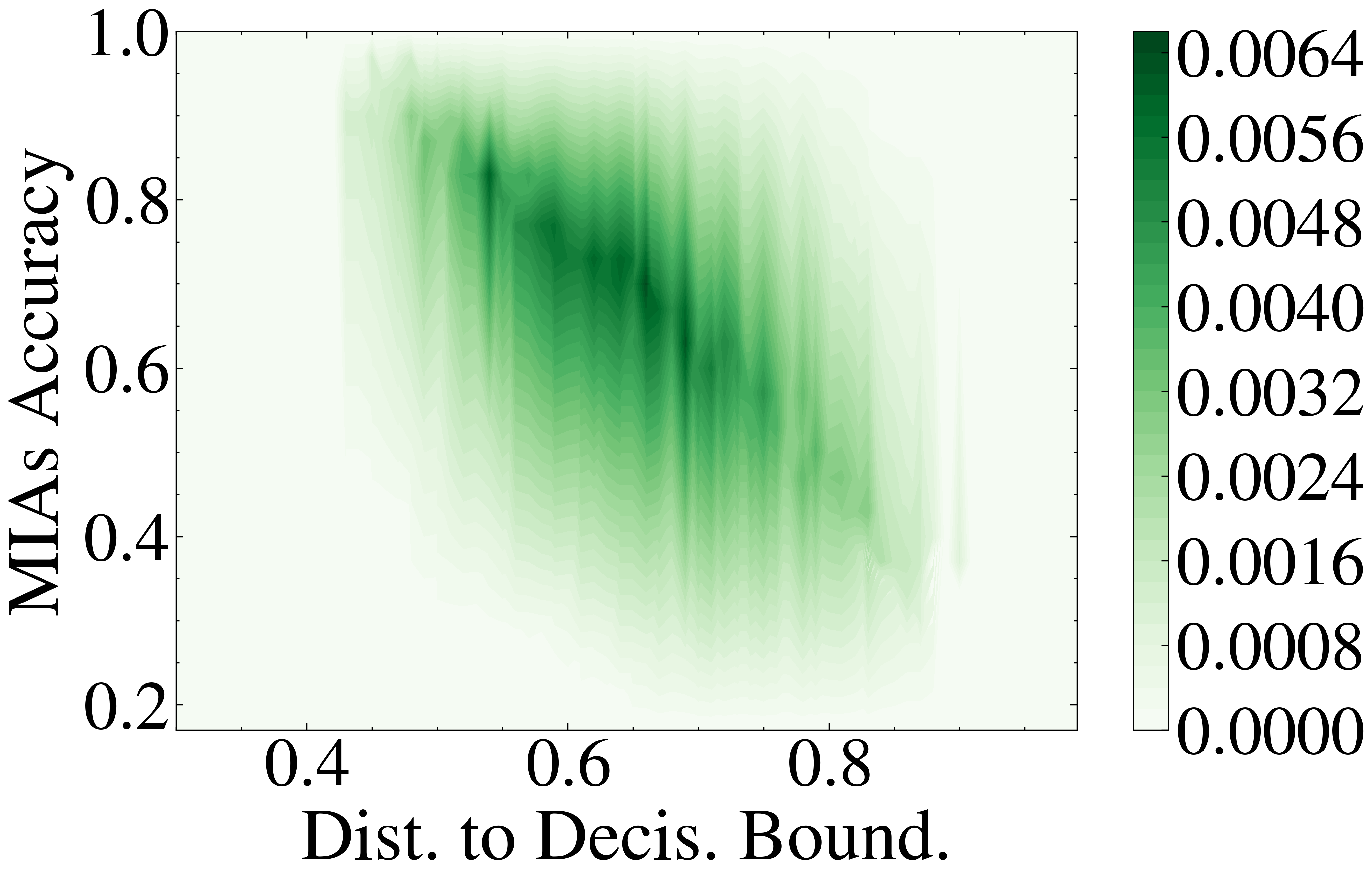}
   \caption{Non-Member (test)}
   \label{fig:d2db_mia_test}
\end{subfigure}
\\
\begin{subfigure}[b]{0.35\textwidth}
         \centering
   \includegraphics[width=\textwidth]{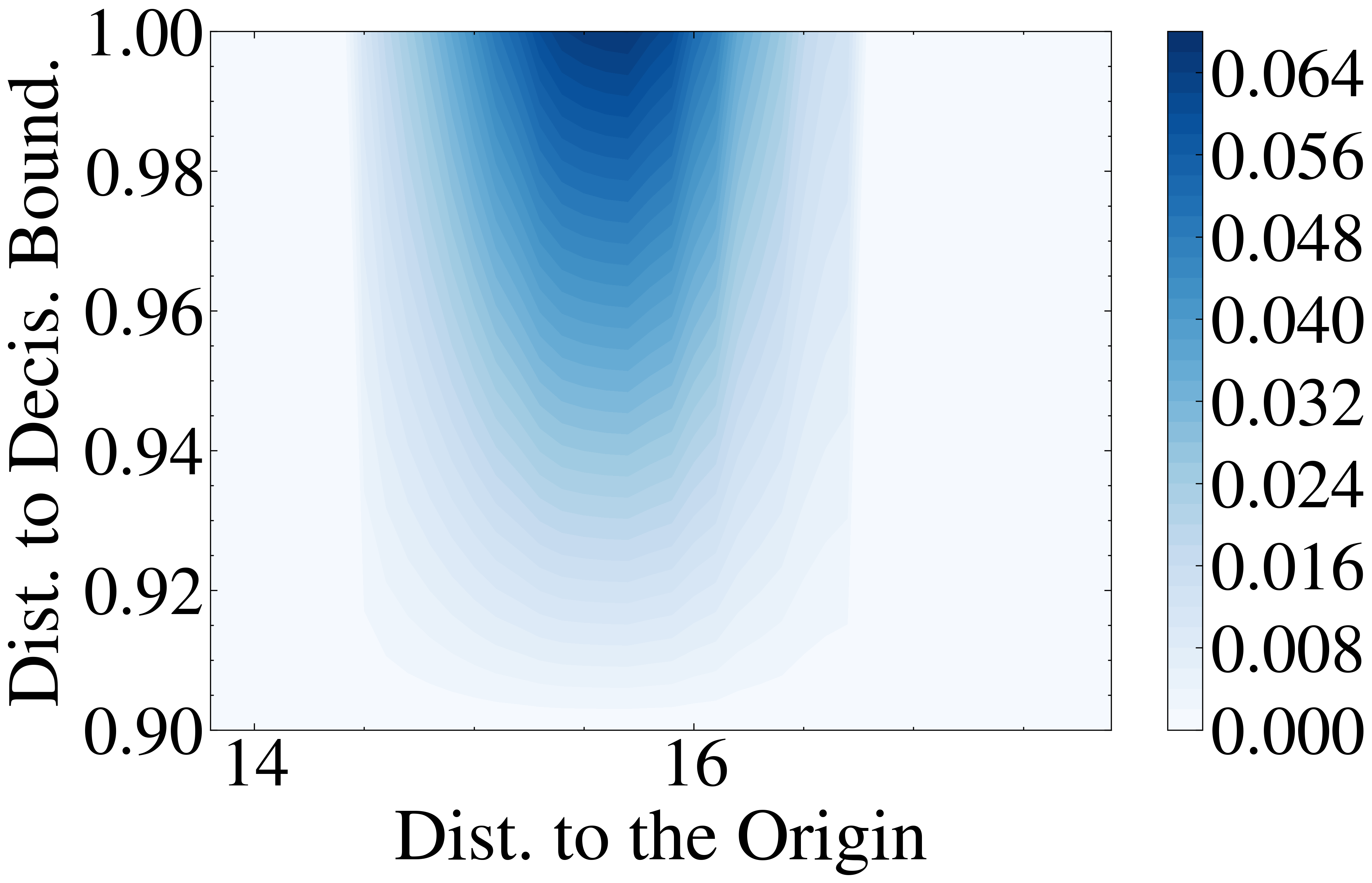}
   \caption{Member (train)}
   \label{fig:d2o_d2db_train}
\end{subfigure}
\quad
\begin{subfigure}[b]{0.35\textwidth}
         \centering
   \includegraphics[width=\textwidth]{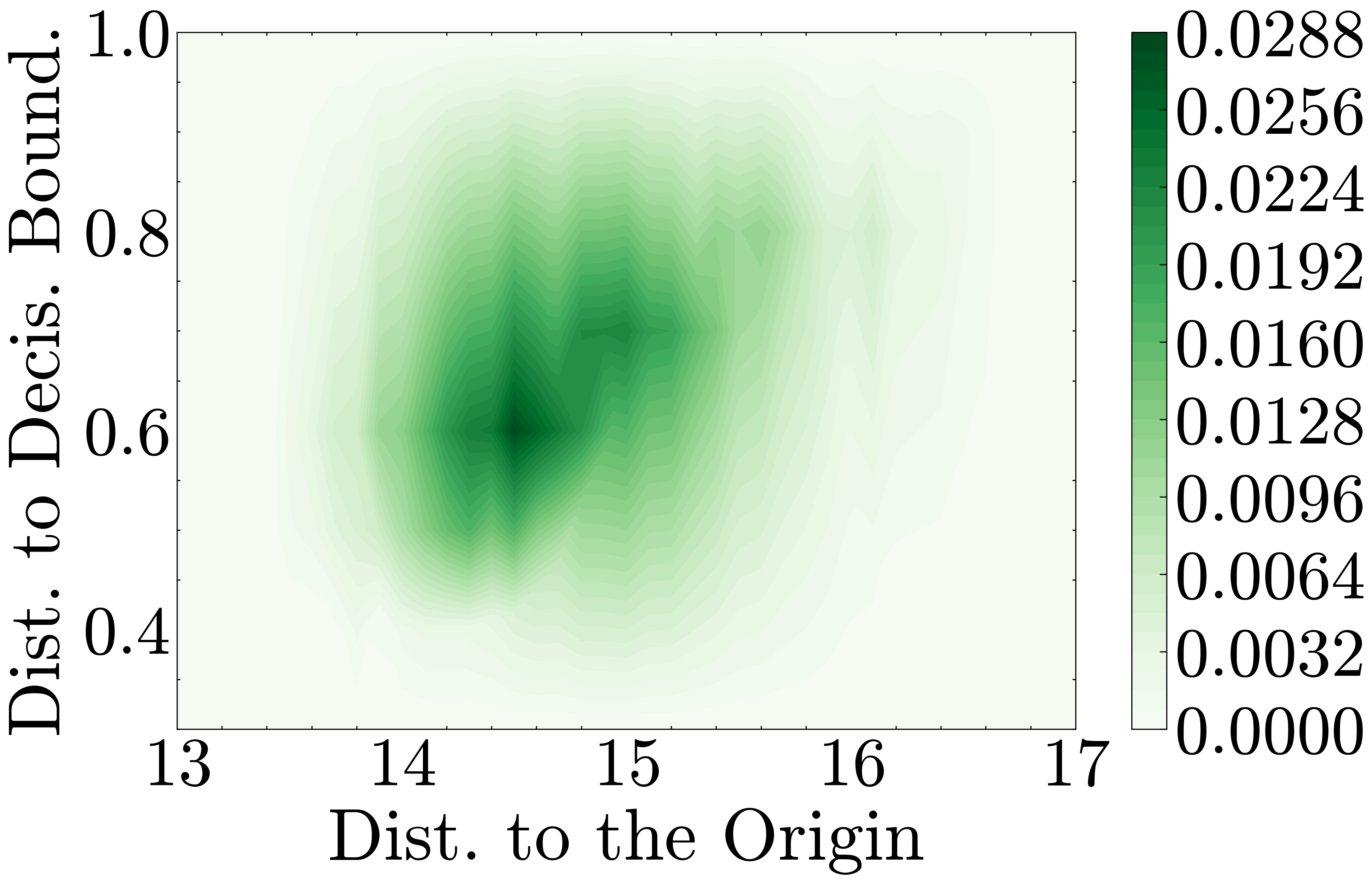}
   \caption{Non-Member (test)}
   \label{fig:d2o_d2db_test}
\end{subfigure}
\end{center}
\caption{\textbf{[1st row]}: the distance to the decision boundary and MIAs accuracy; \textbf{[2nd row]}: the distance to the origin and the distance to the decision boundary. For a sample's distance to the decision boundary, we use the difference between 1st and 2nd maximum prediction probabilities. The results are obtained from dozens of independent experiments. The blue charts ((a) \& (c)) are from train set, and the green charts ((b) \& (d)) are from test set. (ResNet18, CIFAR-100). This figure is excerpted from \protect\cite{fang2024representation}.}
\label{fig:d2o_d2db_mia}
\end{figure}

\subsection{Trade-Offs Between Privacy Preservation and Generalizability}

The behavioral inconsistency of deep learning models in training and testing time, i.e., bad generalizability, leads to privacy-leakage problems. The attacker can steal various information from highly valued samples used to train the model according to this inconsistency.

To show the relationship between representation inconsistency and MIAs accuracy, we visualize the sample-level distribution of the training and testing sets. Fig.~\ref{fig:d2o_d2db_mia} displays the sample-level predictions of MIAs accuracy versus distance to the decision boundary, as well as the relationship between distance to the origin and distance to the decision boundary. The distance to the decision boundary and the distance to the origin are computed from the last and the penultimate layers, respectively. When trained with the standard cross-entropy loss, the model exhibits distinct prediction and attack distributions for members and non-members in both of the layers, indicating that there are multiple privacy-risky layers in the model due to disagreement of representation alignment.

\begin{figure}[t]
     \centering
         \includegraphics[width=0.9\linewidth]{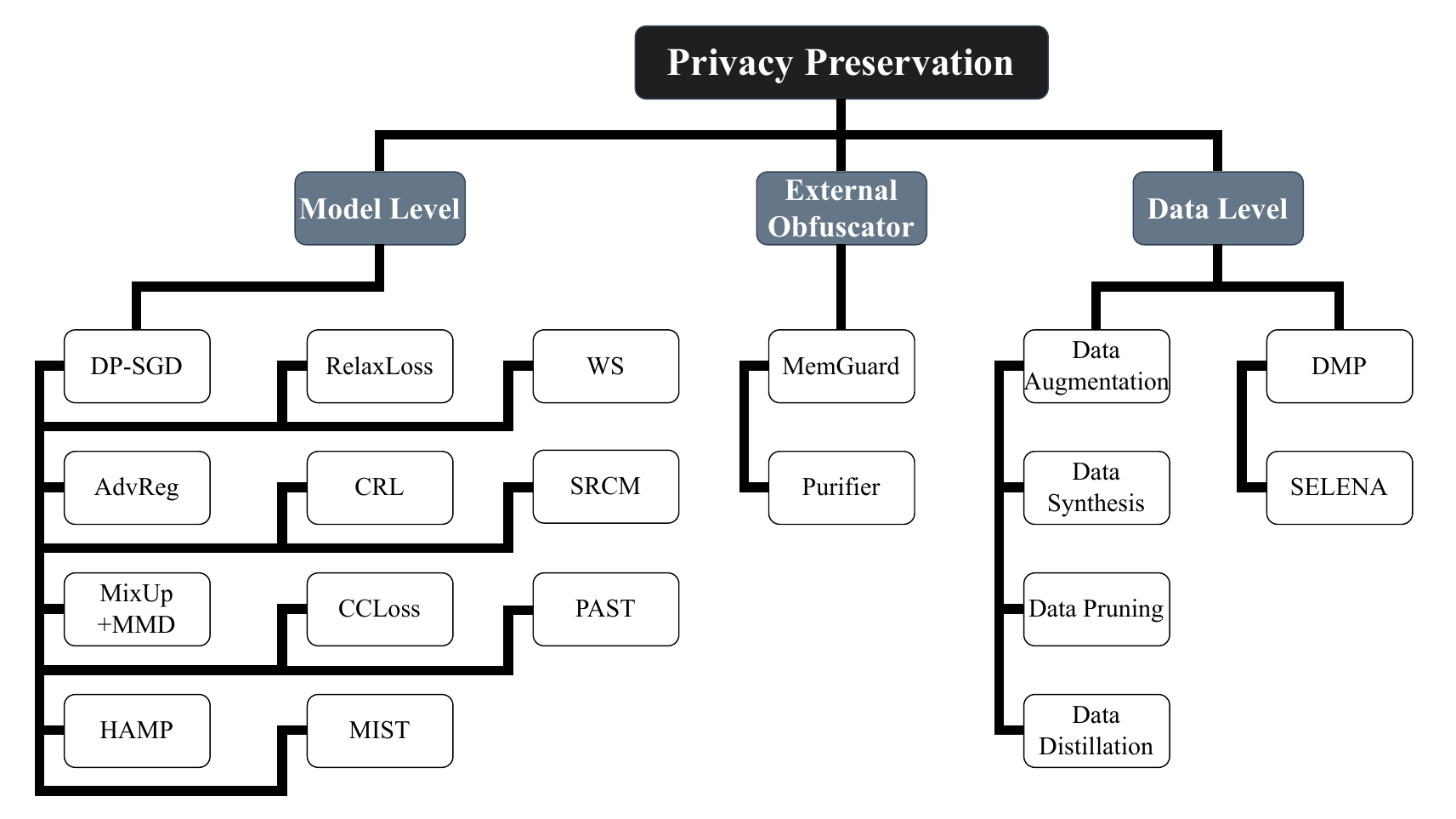}
      \caption{The overview of the privacy preservation approaches.}
  \label{fig:overview_privacy_defence}
\end{figure}

Hence, a straightforward way to mitigate privacy vulnerability is to align the predictions (and representations) between training and testing sets. In the following paragraphs, the introduced approaches try to achieve this alignment goal from different aspects.
In this section, we categorize them into three categories: the model-level solutions, the external obfuscators, and the data-level solutions. The approaches are overviewed in Fig.~\ref{fig:overview_privacy_defence}.

\subsubsection{Model-Level Solutions}
The model-level solutions aim to develop a mechanism to make the prediction distributions aligned with the model's end. 
A classical model-level solution is differentially private stochastic gradient descent (\texttt{DP-SGD}) \cite{abadi2016dpsgd}. It adds noise into the optimizer to prevent the model from taking the (undesirable) easiest way to fit the training data points and also memorizing them.
\cite{nasr2018advreg} introduced an adversarial training framework (\texttt{AdvReg}) that mitigates membership inference attacks by aligning prediction distributions. It tries to develop a discriminator, similar to GAN, to identify the prediction inconsistency of the model while it makes the model try to deceive the discriminator to achieve prediction alignment.
\cite{li2021mixupmmd} (\texttt{MixUp+MMD}) further improved the defending ability by combining mixup data augmentation and maximum mean discrepancy based regularization. 
\cite{chen2022relaxloss} (\texttt{RelaxLoss}) established a threshold to prevent improper fitting for the alignment between member and non-member distributions while preserving the model's generalizability through a technique similar to label smoothing.
\cite{tan2023ws} proposed weighted smoothing (\texttt{WS}) to mitigate memorization by adding normalized random noise to the weights.
Combining the advantages of label-smoothing and \texttt{MemGuard}, \cite{chen2023overconfidence} proposed \texttt{HAMP}, a privacy-beneficial solution on both training and inference stages. 
\cite{liu2024ccloss} incorporated a concave term called Convex-Concave Loss (\texttt{CCLoss}) to lessen the convexity of loss functions, aiming to enhance privacy preservation.
Besides end-to-end solutions, there are also some studies exploring finer-grained solutions.
\cite{fang2024representation} introduced the Saturn Ring Classification Module (\texttt{SRCM}) to bound the representation magnitude to mitigate prediction disparity.
\cite{fang2024crl} tried to align representations in multiple layers by Center-based Relaxed Learning (\texttt{CRL}).
\cite{hu2024past} proposed Privacy-Aware Sparsity Tuning (\texttt{PAST}) to measure weight-level privacy sensitivity and deactivate privacy-risky weights via regularization. 
\cite{li2024mist} (\texttt{MIST}) tried to remove the model's privacy-risky bias via ensembling alignment.

\subsubsection{External Obfuscator}
The external obfuscator is a special kind of model-level solution. Instead of developing a privacy-safe model, it aims to build an obfuscator, which reproduces the prediction probabilities, to remedy the inconsistency in the prediction probabilities.
Similar to the idea of \texttt{DP-SGD}, \texttt{MemGuard} \cite{jia2019memguard} interferes with the prediction confidence distribution of the model by adding additional noise after the model has been trained. 
\cite{yang2023purifier} tried to develop a VAE-based external prediction obfuscator named \texttt{Purifier} to align the prediction probabilities' disparity. Different from \texttt{MemGuard}, it tries to reconstruct the prediction confidences to remove the prediction inconsistency instead of noise confusion.

\subsubsection{Data-Level Solutions}
The data level approaches have two principles: training the privacy-safe model via \textbf{(i) privacy-safe data} or \textbf{(ii) privacy-safe labels}. It is straightforward that when all training data points are privacy-safe, there are no privacy-risky features included in the data, such as shortcut features \cite{geirhos2020shortcut}.

\vspace{0.2in}
\paragraph{Privacy-Safe Data}
The most straightforward solution to produce privacy-safer data is \texttt{Data Augmentation}. There are some data augmentation techniques, such as random cropping and flipping, determined that are able to produce privacy-safer data \cite{kaya2021whendataaug,yu2021howdoesdataaug}. With augmented data, the model can usually achieve better privacy and generalizability. However, there are still no quantifiable metrics to measure how to further produce privacy-safe data through data augmentation. In other words, although the machine learning model can obtain privacy for free via data augmentation, it is unclear if the model achieves complete privacy safety yet.
\cite{stadler2022groundhogday} tried to analyze the effect of synthetic data on the model's privacy (\texttt{Data Synthesis}). 
Besides these two kinds of solutions, there is also an intuitive way to mitigate privacy risks. The first one is \texttt{Data Pruning}. With data pruning techniques, the model can use only a small amount of training data points to develop a well-generalized model. In other words, most membership privacy of the entire train set can still be protected. \cite{ye2024loo_distinguishability} (\texttt{LOOD}) is such a study identifying the privacy-risky data based on Leave-One-Out mechanism. However, the privacy risk mitigation by data pruning is still not perfect, because some membership information of pruned data points could be leaked \cite{li2024datalineageinferenceuncovering}. The other one is \texttt{Data Distillation}. The data distillation aims to refine generalizability-critical features to produce some representative synthetic data. In this process, privacy-risky features can be removed \cite{dong2022privacy_dataset_condensation}. As this direction has not been extensively researched yet, it is foreseen that more studies will be contributed to this topic very soon.

\paragraph{Privacy-Safe Labeling}
Since the prediction disparity in training and testing is due to the improper fitting of the training data points, another way is to stop the model from further fitting into the data when it has learned enough information from the data. This means that if an ideal set of labels exists, the model can be trained perfectly privacy-safe on these labeled data. An intuitive idea is a distillation approach for membership privacy (\texttt{DMP}) \cite{shejwalkar2021dmp}. It trains a protected model via non-member data and produces labels from an unprotected model.
Another solution to better utilize limited data is self-ensemble architecture (\texttt{SELENA}) \cite{tang2022selena}. It developed an ensemble with an efficient sampling strategy to produce privacy-safe labels with better generalizability.

\section{Conclusion}

In this paper, we provide a comprehensive review with regard to Large language model's safety, spurious correlations of deep learning, and privacy, especially membership inference attacks, covering from landmark papers to very recent important literature. 
We believe this paper is a good guide for researchers and practitioners who would like to obtain a good grip on the breadth of the current status of Trustworthy AI research and the depth of particular future agenda.


\bibliography{refs}

\end{document}